How Unlucky is 25-Sigma?
By
Kevin Dowd, John Cotter, Chris Humphrey and Margaret Woods♣

March 24 2008

## 1. Introduction

One of the more memorable moments of last summer's credit crunch came when the CFO of Goldman Sachs, David Viniar, announced in August that Goldman's flagship GEO hedge fund had lost 27% of its value since the start of the year. As Mr. Viniar explained, "We were seeing things that were 25-standard deviation moves, several days in a row."[1] One commentator wryly noted:

> That Viniar. What a comic. According to Goldman's mathematical models, August, Year of Our Lord 2007, was a very special month. Things were happening that were only supposed to happen once in every 100,000 years. Either that … or Goldman's models were wrong (Bonner, 2007b).

But sadly Goldman were not alone. In 2007 alone, massive losses were announced by Bear Stearns, UBS, Merrill Lynch and Citigroup, and then there were the earlier financial disasters – 1987, Daiwa, Barings, Long-Term Capital, the dotcoms, Russia, East Asia, and so on – and afterwards Société Générale and Bear Stearns again in early 2008, with rumours of more yet to come.

Citi's case was particularly interesting. To quote from the same commentator:

> Gary Crittenden, Citi's chief financial officer, claimed … that the firm was simply a victim of unforeseen events. … No mention was made of the previous five years, when Citi was busily consolidating mortgage debt from people who weren't going to repay … pronouncing it 'investment grade' … mongering it to its clients … and stuffing it into its own portfolio … while paying itself billions in fees and bonuses. No, according to the masters of the universe, downgrades by Moody's and Fitch's were completely unexpected … like the eruption of Vesuvius; even the gods were caught off guard. Apparently, as of September 30th, Citigroup's subprime portfolio was worth every penny of the $55 billion that Citi's models said it was worth. Then, whoa, in came one of those 25-sigma events. Citi was whacked by a once-in-a-blue-moon fat tail.
>
> Who could have seen that coming? (Bonner, 2007c).

---

♣ Dowd and Woods: Centre for Risk and Insurance Studies, Nottingham University Business School, Jubilee Campus, Nottingham NG8 1BB, UK. Cotter: Centre for Financial Markets, School of Business, University College Dublin, Carysfort Avenue, Blackrock, Co. Dublin, Ireland. Humphrey: School of Accounting and Finance, University of Manchester, Crawford House, Oxford Road, Manchester M13 9PL, UK. Corresponding author: Kevin.Dowd@nottingham.ac.uk.
[1] Reported in the *Financial Times*, August 13, 2007.

Be all this as it may, one thing is for sure: there are certainly a lot of very unlucky financial institutions around.

**2. How Unlikely is a 25 sigma event?**

The once-in-a-100,000 year figure was quoted in a number of places,[2] and suggests that Goldman, Citi and so on must have been *very* unlucky indeed. But *exactly* how unlikely is a 25-sigma shock?[3]

To start with, lets assume that losses are normally distributed – assume that losses obey the classic bell curve – and ask the question: what is the probability of a loss that is, say, 2 standard deviations or more away from the mean, i.e., what is the probability of a 2-sigma loss event?

The answer is given in Figure 1: the probability associated with a 2 sigma event is equal to the mass of the right-hand tail of the distribution demarcated at the point where the number of standard deviations from the mean is equal to 2, and this is equal to 2.275%.[4] We might therefore expect to see a 2-sigma loss event on one trading day out of 1/2.275%=43.956 days, i.e., on approximately 1 day out of 44 days. A 2-sigma event is unlikely to occur on any given day, but we would expect to see a few of them in any given year.[5]

**Figure: Probability of a 2-Sigma Event**

---

[2] See, e.g., Bonner (2007b,c) Randomwalk (2007) or Hutchinson (2008). A more colourful interpretation, though sadly one that we are unable to verify – was that a 25-sigma event was as likely as Hell freezing over (Bonner, 2007a). Another commentator, Seth Jayson, suggested that a 25-sigma event is as likely as catching an asteroid in the hand (Jason, 2007). Mr. Jayson also made another entertaining analogy, but one that we are sadly unable to repeat.

[3] A straw poll of students and colleagues at Nottingham University Business School carried out by one of the authors indicated that not a single person had the remotest idea of the true probability of a 25 sigma event, and most seemed to think it was likely to occur more often than once in 100,000 years.

[4] For readers who wish to verify it, this number can be obtained using the command '=1-NORMSDIST(2)' in Excel.

[5] Again, in the interests of verifiability: if we go by the common market convention of there being 250 trading days in a year, then we would expect to see about 250/44 or approximately 5.68 such events in any given year.

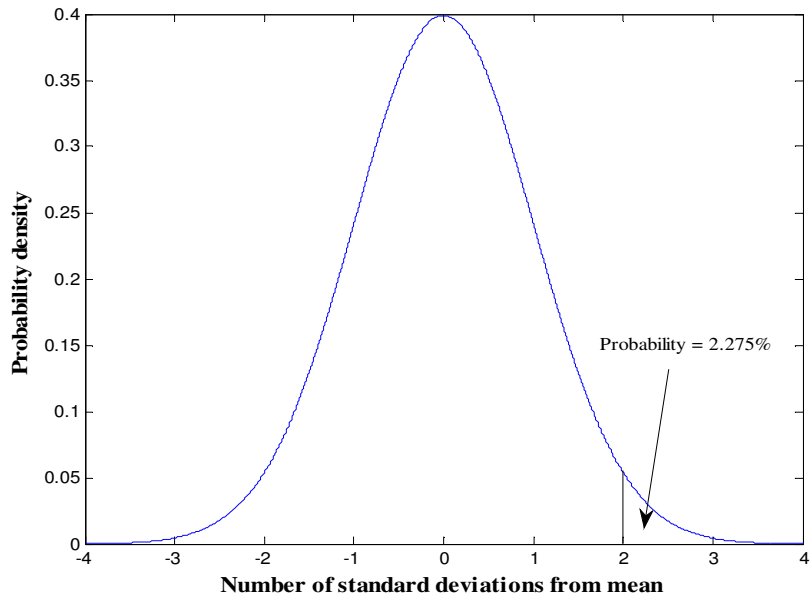

We now use the same approach to estimate the probabilities of losses that are 3 sigmas away from the mean, 4 sigmas away from the mean, and so on. The results of this exercise for 3, 4, 5, 6 and 7 sigma events are given in Table 1.

**Table 1: Probabilities of $k$-sigma events: $k$=3, 4, 5, 6 and 7**

| $k$ | Probability in any given day | Expected occurrence: once in every | |
|---|---|---|---|
| 3 | 0.135% | 740.8 | days |
| 4 | 0.00317% | 31559.6 | days |
| 5 | 0.000029% | 3,483,046.3 | days |
| 6 | 0.000000099% | 1,009,976,678 | days |
| 7 | 0.000000000129% | 7.76e+11 | days |

The reader will note that as $k$ gets bigger the probabilities of a $k$-sigma event fall extremely rapidly:
- a 3-sigma event is to be expected about every 741 days or about 1 trading day in every three years;
- a 4-sigma event is to be expected about every 31,560 days or about 1 trading day in 126 years (!);
- a 5-sigma event is to be expected every 3,483,046 days or about 1 day every 13,932 years(!!)
- a 6-sigma event is to be expected every 1,009,976,678 days or about 1 day every 4,039,906 years;
- a 7-sigma event is to be expected every 7.76e+11 days – the number of zero digits is so large that Excel now reports the number of days using scientific notation, and this number is to be interpreted as 7.76 days with decimal point pushed back 11 places. This frequency corresponds to 1 day in 3,105,395,365 years.

These results are breathtaking. To give them some perspective: a 5-sigma event corresponds to an expected occurrence of less than just one day in the entire period since the end of the last Ice Age; a 6-sigma event corresponds to an expected occurrence of less than one day in the entire period since our species, Homo Sapiens, evolved from earlier primates; and a 7-sigma event corresponds to an expected occurrence of just once in a period approximately five times the length of time that has elapsed since multicellular life first evolved on this planet.[6]

At this point, there are so many decimal points in the numbers involved that Excel is unable to handle values of $k$ bigger than 7, and we are still a very long way from the 25 sigma events that Goldman and Co. experienced. To get around the limitations of Excel, we now switch over to MATLAB and use the MATLAB command '1-normcdf(8,0,1)' to estimate the corresponding probability for an 8-sigma event, and this corresponds to an expected occurrence once every 6.429e+012 years. To put this into perspective, this is a period (considerably) longer than the entire period that has elapsed since Big Bang.[7] If we observe a profit or loss once a day, then a mere 8-sigma event should occur less than once in the entire history of the universe.

We then moved to 9 sigma events, but ran into a problem because the MATLAB 'normcdf' function gives 9 sigma and larger events a flat value of zero: the probabilities are so small they now fall under the function's radar. To get around this problem we wrote a specially designed MATLAB function to estimate the probabilities and expected occurrence periods associated with larger losses, and the details involved are reported in the Appendix. We then used this function to produce the high-sigma results shown in Table 2:

**Table 2: Probabilities of High Sigma Events**

| $k$ | Probability in any given day | Expected occurrence: once in every | |
|---|---|---|---|
| 10 | 7.620e-022 % | 5.249e+020 | years |
| 15 | 3.671e-049 % | 1.090e+048 | years |
| 20 | 2.754e-087 % | 1.453e+086 | years |
| 25 | 3.0570e-136 % | 1.309e+135 | years |

These numbers are on truly cosmological scales, and a natural comparison is with the number of particles in the Universe, which is believed to be between 1.0e+73 and 1.0e+85 (Clair, 2001). Thus, a 20-event corresponds to an expected occurrence period measured in years that is 10 times larger than the higher of the estimates of the number of particles in the Universe. For its part, a 25-sigma event corresponds to an expected occurrence period that is equal to the higher of these estimates but with the decimal point moved 52 places to the left!

To give a more down to earth comparison, on February 29 2008, the UK National Lottery is currently was offering a prize of £2.5m for a ticket costing £1. Assuming it to be a fair bet, the probability of winning the lottery on any given attempt is therefore 0.0000004. The probability of winning the lottery $n$ times in a row is therefore $0.0000004^n$, and the probability of a 25 sigma event is comparable to the probability of winning the lottery 21 or 22 times in a row.[8]

---

[6] The end of the last Ice Age was about 10,000 years ago; Homo Sapiens evolved within the last million years; and multicellular life on Earth originated about 600 million years ago.
[7] The NASA website reports that Big Bang occurred between 12 and 14 billion years ago.

And we should not forget Goldman's losing streak – Goldman did not just experience a single 25-sigma event, but experienced several in a row – or forget that other institutions also experienced 25-sigma events. If the probability of a single 25-sigma event is low, the odds of two or more such events are truly infinitesimal. For example, the odds of two 25-sigma events on consecutive days are equal to 3.057e-136 squared, which is 9.3450e-272. This is as likely as winning the lottery about 42 times in a row. The corresponding expected occurrence period is the square of 1.309e+135 years – that is, 1.713e+270 years – a number so vast that it dwarves even cosmological figures. As Oscar Wild might have put it: to experience a single 25-sigma event might be regarded as a misfortune, but to experience more than one does look like carelessness.

It is pretty clear by now that a 25-sigma event is much, much, *much* less likely than is suggested by an expected occurrence every 100,000 years. In fact, if we take the true figure (1.309e+135) and divide it by 100,000 we get a figure equal to 1.309e+130, and this tells us that the estimate of 100,000 is out by 130 decimal points: it is not even close to within 100 decimal points. We suspect, on the other hand, that the estimate of a 25-sigma event being on a par with Hell freezing over is probably about right.

**3. Bad luck or just incompetence?**

However low the probabilities, and however frequently 25-sigma or similar events actually occur, it is always *possible* that Goldman's and other institutions that experienced such losses were just unlucky – albeit to an extent that strains credibility.[9] But if these institutions are really *that* unlucky, then perhaps they shouldn't be in the business of minding other people's money. Of course, those who are more cynical than us might suggest an alternative explanation – namely, that Goldmans and their Ilk are simply not competent at their job. Heaven forbid!

All this poses an interesting dilemma for investors: would you prefer the people looking after your money to be incredibly unlucky or just plain incompetent? There again, maybe there is truth in both explanations. Benjamin Franklin once aptly observed that "Diligence is the mother of good luck", and it would stand to reason that their opposites, bad luck and incompetence, might also be related.

As for Mr. Viniar: he promised that Goldman would take a more "robust approach" in future, and this will no doubt be welcome news to hard-hit investors and the millions facing foreclosure on their mortgages. He went on to acknowledge that the experience "makes you reassess how big the extreme moves can be".[10] Indeed.

---

[8] Note that $0.0000004^{21} = 4.3980\text{e-}135$ and $0.0000004^{22} = 1.7592\text{e-}141$ so the probability of a 25-sigma event lies somewhere between the two.

[9] Defenders of the 'bad luck' scenario might argue that the probabilities are only as low as they are because we have assumed normality, whereas real world loss distributions are fat-tailed and fat-tailed distributions lead to much higher probabilities of extreme events. This is true but irrelevant: so what if the true probability of a 25-sigma event is 3.057e-36 rather than 3.057e-136? Moving the decimal point a hundred places or more makes no practical difference. We would invite anyone who disagrees to do the calculations for themselves.

Funny things these 25-sigma events. And surprisingly common too.

**Appendix: Estimating the Probabilities of High Sigma Events**

Following Zeghbroeck (2007), let the Gaussian probability density be $G(x)$ and the error function be $erf(x)$. These are related via

(1) $$\int_{-x}^{x} G(x)dx = erf\left(\frac{x}{\sigma\sqrt{2}}\right)$$

If we set $\sigma = 1$ then $x$ is the number of standard deviations away from the mean. Hence,

(2) $$\int_{-x}^{x} G(x)dx = erf\left(\frac{x}{\sqrt{2}}\right)$$

---
[10] *Financial Times*, August 13, 2007.

Let $p$ be the tail probability associated with some value $x$. It follows from rearranging (2) that

(3) $$p = 0.5 - 0.5 \times erf\left(\frac{x}{\sqrt{2}}\right)$$

However, it is also known that for large $y$,

(4) $$erf(y) \approx 1 - \frac{e^{-y^2}}{y\sqrt{\pi}}\left(1 - \frac{1}{2y^2} + \frac{1 \times 3}{(2y^2)^2} + \frac{1 \times 3 \times 5}{(2y^2)^3} + ...\right)$$

It then follows from further rearranging that

(5) $$p \approx \frac{e^{-y^2}}{2y\sqrt{\pi}}\left(1 - \frac{1}{2y^2} + \frac{1 \times 3}{(2y^2)^2} + \frac{1 \times 3 \times 5}{(2y^2)^3}\right)$$

where $y = x/\sqrt{2}$.

The MATLAB function used to produce the high-sigma results reported in the paper is reproduced below:

```
function p=high_sigma_prob(x)
% Function estimates p, the probability of a high-sigma loss event.
%
% Input:
%  - x: size of loss in terms of number of sigmas away from mean
%
% Written by Kevin Dowd March 1, 2008.
%*******************************************************************
y=x/sqrt(2);
p=0.5*(exp(-y^2)/(sqrt(pi)*y))*...
(1-1/(2*y^2)+1*3/(2*y^2)^2+1*3*5/(2*y^2)^3);
```